\documentclass[10pt,aps,prl,twocolumn,preprintnumbers,nofootinbib]{revtex4-2}
\usepackage{amsmath,amssymb,mathtools}
\usepackage{hyperref}
\usepackage{xspace}
\usepackage{graphicx}
\usepackage{grffile}
\usepackage{enumitem}
\usepackage{xcolor}

\newcommand{\der}{\mathrm{d}}

\newcommand{\Fcal}{\mathcal{F}}
 \newcommand{\Ccal}{\mathcal{C}}
\newcommand{\xh}{\hat{x}}
\newcommand{\zh}{\hat{z}}

\newcommand{\code}[1]{\texttt{#1}}

\newcommand{\dd}{\mathrm{d}}

\begin{document}

\preprint{ZU-TH 25/25}

\title{Identified Hadron Production in Deeply Inelastic Neutrino-Nucleon Scattering}%

\author{Leonardo Bonino}
\affiliation{%
  Physik-Institut, Universit\"at Z\"urich, Winterthurerstrasse 190, 8057 Z\"urich, Switzerland}%
\author{Thomas Gehrmann}
 \affiliation{%
  Physik-Institut, Universit\"at Z\"urich, Winterthurerstrasse 190, 8057 Z\"urich, Switzerland}%
\author{Markus L\"ochner}
 \affiliation{%
  Physik-Institut, Universit\"at Z\"urich, Winterthurerstrasse 190, 8057 Z\"urich, Switzerland}%
\author{Kay Sch\"onwald}
 \affiliation{%
  Physik-Institut, Universit\"at Z\"urich, Winterthurerstrasse 190, 8057 Z\"urich, Switzerland}%
\author{Giovanni Stagnitto}
\affiliation{%
Universit\`{a} degli Studi di Milano-Bicocca \& INFN, Piazza della Scienza 3, 20216 Milano, Italy}%

%\date{\today}

\begin{abstract}
  The production of identified hadrons in semi-inclusive deep-inelastic scattering (SIDIS) is sensitive to
  parton distribution functions and hadron fragmentation functions. Neutrino-induced
  SIDIS processes probe combinations of these functions different from their
  charged-lepton-induced counterparts.  We compute charged pion production
 in  \mbox{(anti-)}neutrino induced SIDIS up to second order in perturbative QCD
 and compare our predictions to precise legacy fixed-target data. We demonstrate
 the high sensitivity of these data on the parametrization of the fragmentation functions and discuss
 future SIDIS probes at the LHC Forward Physics Facility.
\end{abstract}

\maketitle

\textit{Introduction---}\ignorespaces
Neutrino beams are ideal probes for detailed studies of the partonic content of
nucleons in deep-inelastic scattering (DIS) experiments.
Of particular interest are \mbox{(anti-)}neutrino-nucleon scattering processes with an identified, highly energetic secondary muon.
In these processes the momentum transfer between \mbox{(anti-)}neutrino and nucleon is mediated by virtual $W$ bosons, resulting in a flavor-changing charged current (CC) interaction on the quark line.
This interaction singles out specific  flavor combinations
of the parton distribution functions (PDFs), providing complementary information with respect to neutral current (NC) DIS.

The flavor-discriminating power of CC processes is further increased by additionally identifying a final state hadron, constraining the allowed flavor combinations of initial and final state quarks.
The identification of final state hadrons also provides crucial information on the mechanism of hadronization, responsible for the formation of color neutral hadronic states out of quarks and gluons.
CC DIS data extracted in neutrino experiments \cite{Berge:1989hr,CCFRNuTeV:2000qwc,CHORUS:2005cpn,NOMAD:2013hbk} are routinely included in global PDF fits~\cite{Hou:2019efy,Bailey:2020ooq,NNPDF:2021njg}, which are being performed up to next-to-next-to-leading
order (NNLO) in QCD.
As of today no global fits to light hadron fragmentation functions (FFs) include CC
neutrino-nucleon SIDIS information,
despite this process providing a clean probe of the quark flavor structure of FFs, which is otherwise
poorly constrained from other fragmentation observables.

In this Letter we present precise predictions for semi-inclusive pion production in \mbox{(anti-)}neutrino-nucleon scattering at NNLO in massless QCD.
We compare our predictions with data from the Aachen-Bonn-CERN-Munich-Oxford (ABCMO) collaboration~\cite{Aachen-Bonn-CERN-Munich-Oxford:1982jrr}, which has never been included in any modern extraction of PDFs or FFs.
We demonstrate the potential impact of this data set on future determinations of FFs and discuss the
 potential of future neutrino-nucleon SIDIS at the Forward Physics Facility at CERN.

\textit{SIDIS Cross Section---}\ignorespaces
We consider the observation of a hadron $h$ following the scattering of an \mbox{(anti-)}neutrino on a nucleon,
\begin{align}
&\nu(k)\,  p(P) \to l^{-}(k^{\prime})\, h(P_h) \,X \,  , \nonumber \\
& \bar{\nu}(k)\,  p(P) \to l^{+}(k^{\prime})\, h(P_h) \, X \, ,
\end{align}
with $X$ denoting the remaining hadronic final state.
The leptonic momenta determine the four-momentum $q=k-k^{\prime}$
of the exchanged virtual $W$-boson and the rest-frame energy transfer $y=(P\cdot
q)/(P\cdot k)$.   For $Q^2=-q^2$ the variables
\begin{align}
x=\frac{Q^2}{2P\cdot q}\, ,\quad \quad z=\frac{P\cdot P_h}{P\cdot q}
\end{align}
correspond to the Born-level momentum fractions of the
nucleon carried by the incoming parton ($x$) and of the outgoing parton carried by the identified hadron~($z$).
The squared center-of-mass energy of the lepton-nucleon system is $s=Q^2/(xy)$ and
the invariant mass of the hadronic final state $W^2 = Q^2(1-x)/x$.

Following the notation of \cite{deFlorian:2012wk}, the triple-differential cross section reads
\begin{align}\label{eq:d3sigdxdydz}
\frac{\dd^3 \sigma^{h}}{\dd x\,\dd y\,\dd z} \hspace{-1pt} = \hspace{-1pt} \frac{16\pi \alpha^2}{Q^2} \eta_W \hspace{-1pt} & \Bigg[  \frac{1+(1-y)^2}{2y}\mathcal{F}_T^{h, W^{\pm}} \hspace{-2.55pt} + \hspace{-1.5pt} \frac{1-y}{y}\mathcal{F}_L^{h,W^{\pm}}  \nonumber \\
&-e_{\ell} \frac{1-(1-y)^2}{2y}\mathcal{F}_3^{h,W^{\pm}}  \Bigg] \,  ,
\end{align}
where $\alpha$ is the fine structure constant, $e_{\ell}$ refers to the outgoing lepton charge, and
\begin{align}
\eta_W = \frac{1}{2}\left( \frac{G_F M_W^2}{4\pi \alpha}\frac{Q^2}{Q^2+M_W^2}\right)^2 \, .
\end{align}

The charged current (CC) SIDIS structure functions $\Fcal^{h,W^{\pm}}_T$, $\Fcal^{h,W^{\pm}}_L$ and $\Fcal^{h,W^{\pm}}_3$ are obtained by summing over all partonic channels of the convolution between the PDF for a parton $p$ ($f_p$), the FF of a parton $p^{\prime}$ into
the hadron $h$ ($D^h_{p^{\prime}}$), and the coefficient function for the transition $p\to p^{\prime}$ ($\Ccal^{i,k}_{p' p}$):
\begin{align}
\Fcal_i^{h,W^{\pm}}&(x,z,Q^2) = \sum_{p,p'}\int_x^1 \frac{\dd \xh}{\xh} \int_z^1 \frac{\dd \zh}{\zh} \, f_p\bigg(\frac{x}{\xh},\mu_F^2\bigg) \nonumber \\ & D^h_{p'}\bigg(\frac{z}{\zh},\mu_A^2\bigg)
\Ccal^{i,W^{\pm}}_{p'p}(\xh,\zh,Q^2,\mu_R^2,\mu_F^2,\mu_A^2) \,  ,
\end{align}
for $i=T,L,3$. The above factorization introduces two separate
factorization scales: $\mu_F$ for the initial state and $\mu_A$ for the final
state.  $\mu_R$ denotes the renormalization scale.  The coefficient
functions encode the hard-scattering part of the process and can be computed in
perturbative QCD. Their perturbative expansion in the strong coupling constant
$\alpha_s$ reads
\begin{align}
\mathcal{C}^{i,W^{\pm}}_{p'p}&=C^{i,W^{\pm}, (0)}_{p'p}+\frac{\alpha_s(\mu_R^2)}{2\pi}C^{i,W^{\pm}, (1)}_{p'p}\nonumber\\
&+\left(\frac{\alpha_s(\mu_R^2)}{2\pi}\right)^2C^{i,W^{\pm}, (2)}_{p'p} +\mathcal{O}(\alpha_s^3)\,  .
\end{align}
Analytical expressions for the SIDIS coefficient functions have been computed up to next-to-leading order (NLO) in QCD \cite{Altarelli:1979kv,Baier:1979sp,Furmanski:1981cw}.

For this Letter we have performed the first computation of the full set of NNLO corrections to neutrino-induced SIDIS in analytical form~\cite{Bonino:2025qta}, closely following our earlier calculation of photon-mediated SIDIS~\cite{Bonino:2024qbh,Bonino:2024wgg}
at this order. The technical details of our calculation are described in~\cite{Gehrmann:2022cih,Bonino:2024adk},
and an independent validation of~\cite{Bonino:2024qbh,Bonino:2024wgg}
was obtained in~\cite{Goyal:2023zdi,Goyal:2024tmo,Goyal:2024emo,Ahmed:2024owh}.
In order to treat the antisymmetric structures $\gamma_5$ and $\varepsilon^{\mu\nu\rho\sigma}$ arising from axial couplings and the projector of $\Fcal_3^{h,W^{\pm}}$ consistently in dimensional regularization, we employed the Larin scheme~\cite{Larin:1993tq},
and subsequently converted the results into the $\overline{\text{MS}}$ scheme by a finite transformation.

\textit{Numerical Results---}\ignorespaces
The production of charged pions in deeply-inelastic \mbox{(anti-)}neutrino-proton scattering was measured by the ABCMO collaboration~\cite{Aachen-Bonn-CERN-Munich-Oxford:1982jrr} in 1982, alongside results from several other contemporaneous
experiments~\cite{Bell:1978ta,Derrick:1981br,Amsterdam-Bologna-Padua-Pisa-Saclay-Turin:1981hcw,Aachen-Bonn-CERN-Democritos-London-Oxford-Saclay:1981qxq,Amsterdam-Bergen-Bologna-Padua-Pisa-Saclay-Turin:1983mnk,Amsterdam-Bergen-Bologna-Padua-Pisa-Saclay-Turin:1983mnk,AMSTERDAM-BERGEN-BOLOGNA-PADUA-PISA-SACLAY-TURIN:1984stg,WA25:1988mir,Birmingham-CERN-ImperialColl-MunichMPI-Oxford-UnivCollLondon:1991hsx}.
The experiment exploited the wide-band \mbox{(anti-)}neutrino beam generated by 350 GeV and 400 GeV protons from the CERN SPS
and directed onto the liquid hydrogen-filled Big European Bubble Chamber (BEBC).
It measured the SIDIS processes
\begin{align}
&\nu\,  p \to \mu^{-}\, \pi^{\pm} \,X \,  , &
&\bar{\nu}\,  p \to \mu^{+}\, \pi^{\pm} \, X \,  ,
\end{align}
providing separate results for both pion charges.

The experimental data was analyzed in the variable $x_\mathrm{F}$, which differs from $z$ at higher perturbative orders due to the transverse momentum of extra real radiation.
We have performed a detailed analysis of the resulting shift in predictions differential in $x_\mathrm{F}$ and $z$ up to NLO and find that for the kinematics under consideration we can safely approximate $x_\mathrm{F} \simeq z$.

\begin{figure*}[t]
\includegraphics[width=\textwidth]{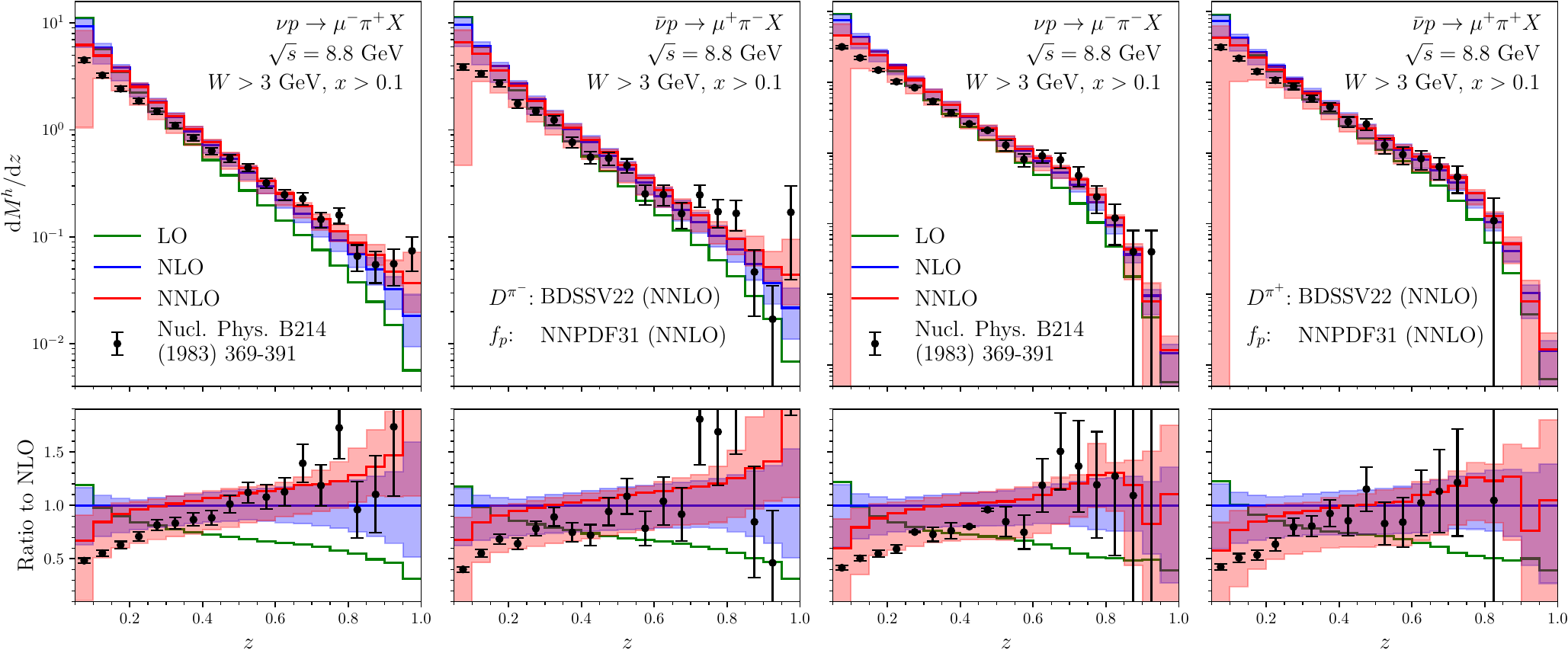}
\caption{Theory predictions for pion multiplicities for (anti-)neutrino induced DIS processes up to NNLO compared to the ABCMO measurement~\cite{Aachen-Bonn-CERN-Munich-Oxford:1982jrr}.
}
\label{fig:ABCMO}
\end{figure*}

The normalized distributions of Table 3 in~\cite{Aachen-Bonn-CERN-Munich-Oxford:1982jrr} correspond to
charged pion multiplicities described by the ratio
\begin{align}\label{eq:RhSIDIS}
\frac{\der M^{\pi^{\pm}}}{\der z} = \frac{\der^3 \sigma^{\pi^{\pm}}/\der x\,\der y \,\der z}{\der^2 \sigma/\der x\,\der y} \, .
\end{align}
The numerator is given by the SIDIS cross section of eq.~\eqref{eq:d3sigdxdydz}, while the denominator is given by the inclusive DIS cross section.
The NNLO DIS structure functions~\cite{Zijlstra:1992qd,Zijlstra:1992kj} are computed using \mbox{\code{APFEL}} \cite{Bertone:2013vaa}.
The measured multiplicities allow to extract FFs.

As the ABCMO~\cite{Aachen-Bonn-CERN-Munich-Oxford:1982jrr} results do not report the average \mbox{(anti-)}neutrino beam energy,  we infer an average beam energy of $E_{\nu/\bar{\nu}}\simeq 39 \,\mathrm{GeV}$ ($\sqrt{s_\mathrm{avg}} \simeq 8.8\, \mathrm{GeV}$) from another group~\cite{Amsterdam-Bologna-Padua-Pisa-Saclay-Turin:1981hcw} using the same experimental setup.
While a weighted average over the beam energy $E_{\nu/\bar{\nu}}$ is in principle required,  the exact distributions are not reported in~\cite{Aachen-Bonn-CERN-Munich-Oxford:1982jrr,Amsterdam-Bologna-Padua-Pisa-Saclay-Turin:1981hcw}.
We instead verified that changing the beam energy between $19.5\,\mathrm{GeV}$ and $78.0\,\mathrm{GeV}$ leads to small variations that are evenly spread around the central value, justifying the assumption of a mono-energetic beam.
The numerator and denominator of eq.~\eqref{eq:RhSIDIS} are integrated over $x$, $y$ and $z$ according to the kinematic cuts reported in~\cite{Aachen-Bonn-CERN-Munich-Oxford:1982jrr},
$x>0.1$ and $W > 3\, \mathrm{GeV}$,
resulting in a dynamical range $1\,\mathrm{GeV} < Q  < 8.8\,\mathrm{GeV}$.

\begin{figure*}[t]
 \includegraphics[width=\textwidth]{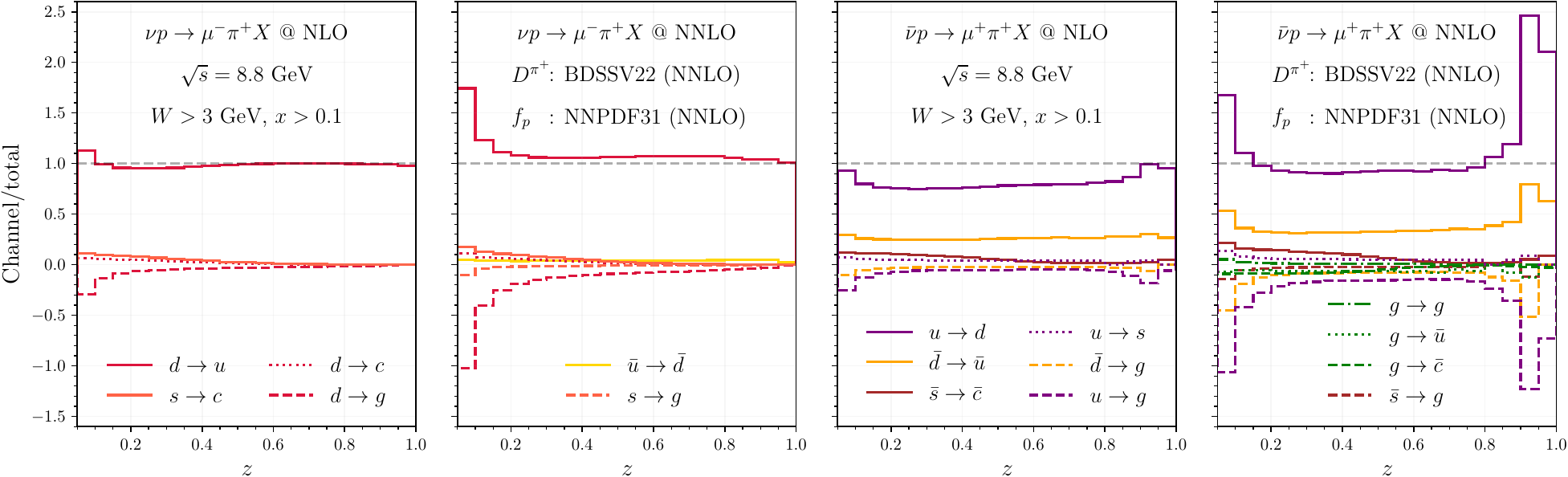}
  \caption{Dominant partonic channels contributing to flavor-favored (left frames) and flavor-unfavored (right frames)
  SIDIS reactions at NLO and NNLO.
Contributions are normalized to unity in each bin. }
  \label{fig:channels}
\end{figure*}

Our results are shown in Fig.~\ref{fig:ABCMO}.  At each perturbative order we use the NNLO PDF set \mbox{\code{NNPDF31\_NNLO}} \cite{NNPDF:2017mvq} and NNLO FF set \mbox{\code{BDSSV22\_NNLO}} \cite{Borsa:2022vvp}. $\alpha_s$ is taken from the PDF set, and the CKM values are from
\cite{ParticleDataGroup:2024cfk}.
We fix the number of flavors to $N_f = 4$ in all ingredients of the calculation and keep all quarks massless.
In inclusive charged current processes, charm quark mass effects may be substantial~\cite{Gottschalk:1980rv, Gluck:1997sj, Behring:2016hpa, Behring:2015roa} due to the $s\to c$ transition at tree level, which is however suppressed by the fragmentation functions for semi-inclusive pion production.

The theory uncertainties are obtained by seven-point scale variation in the numerator for $\mu_R$ and $\mu_A=\mu_F$ around the central scale $Q$, with scale variations cut off at the minimum value $Q_{\rm min}$ allowed by the PDF and FF
sets.
We verified that an additional scale variation in the denominator 
or an uncorrelated variation of $\mu_A$ and $\mu_F$ have a
negligible effect on the theory uncertainty band.

The legacy data shows strikingly good compatibility with the theoretical predictions.
While the LO predicts the overall trend,  the inclusion of the NLO corrections shifts the prediction closer to the data.
The NNLO corrections are sizable and further improve agreement with the data,
now
reproducing the trend of the data in the entire $z$ range, with the majority of the data points lying within theory uncertainties.
For small and moderate $z$ the experimental errors on the data points are significantly smaller than theory uncertainties.
The theoretical uncertainties from scale variation at NLO and NNLO are comparable in size,  hinting at an underestimation of NLO uncertainties.

\begin{figure*}[t]
 \includegraphics[width=\textwidth]{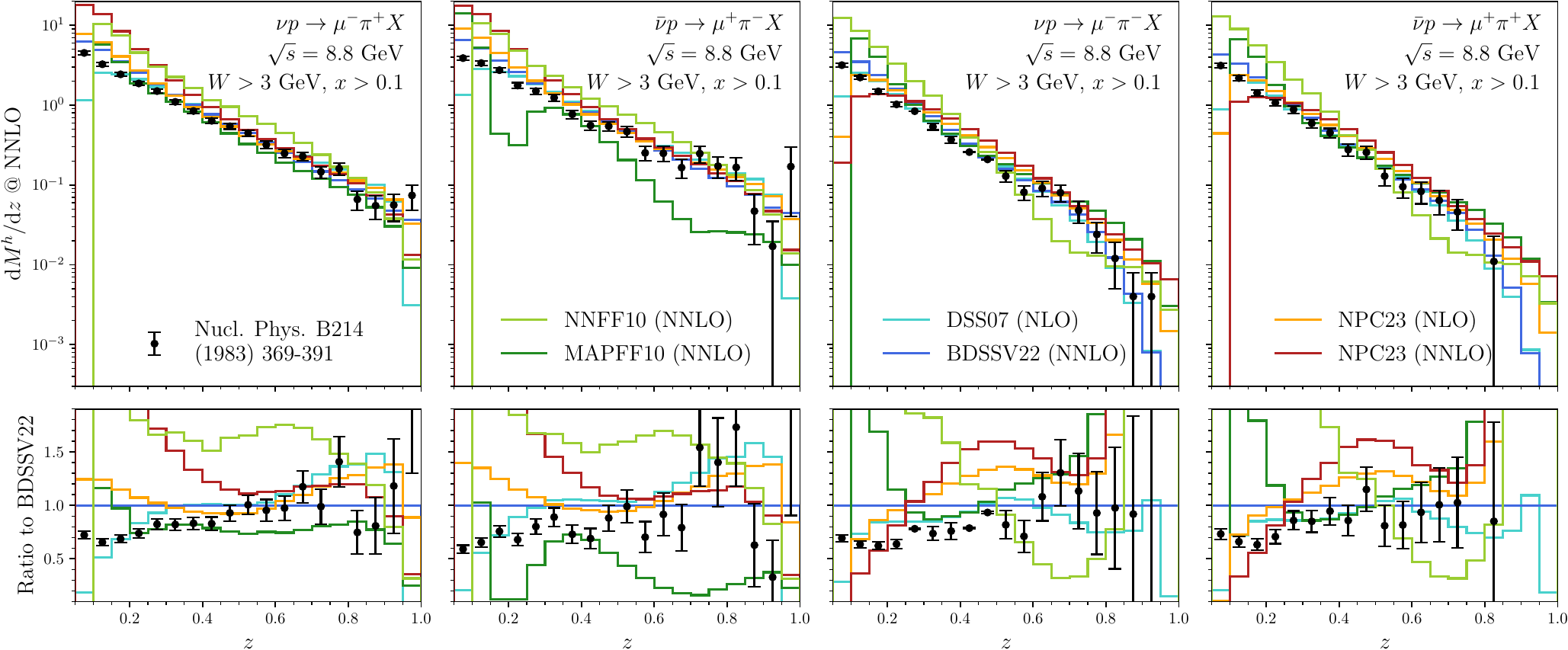}
  \caption{Comparison of NNLO multiplicities computed with different FF sets to data.  Ratios are taken with respect to the multiplicity computed with the \code{BDSSV22\_NNLO} set. }
  \label{fig:ff_comparison}
\end{figure*}

In Fig.~\ref{fig:channels} we illustrate the channel decomposition for $\nu p \to \mu^-\pi^+ X$ and $\bar{\nu} p \to \mu^+\pi^+ X$ at NLO and NNLO for ABCMO kinematics.
Only channels contributing to at least $5 \%$ of the total cross section in a single bin are shown, displaying the
sensitivity of this data on off-diagonal flavor combinations in the underlying hard interaction, which in turn
probe specific FFs.

For $\nu p \to \mu^-\pi^+ X$ the full cross section is essentially accounted for by the flavor combination \emph{favored} by the valence content of the
proton and $\pi^+$: $d \to u$.  This transition is a non-singlet CKM-allowed channel already present at Born level.
For $\bar{\nu} p \to \mu^+\pi^+ X$ the \emph{favored} valence-to-valence flavor combination channels, containing
a $u$-quark in the final state, arise only at NNLO and are thus perturbatively suppressed.
Instead, the \emph{unfavored} sea-flavor combinations are dominant, accounting for a significant fraction of the total SIDIS cross section.

The NNLO corrections even enhance the impact of unfavored contributions compared to NLO,
and they enlarge the negative $u\to g$ and $d\to g$ channels at small $z$, bringing the overall cross section closer to
the data.
The induced large cancellations increase the theoretical uncertainty,
as already observed for charged-lepton-induced SIDIS~\cite{Bonino:2024qbh,Bonino:2024wgg}.
Especially in the high $z$ regions, instabilities in the flavor-unfavored channels at NNLO in Fig.~\ref{fig:ABCMO} are
due to large cancellations among partonic channels.
The related cross sections $\bar{\nu} p \to \mu^+ \pi^- X$ (favored)
and  $\nu p \to \mu^- \pi^- X$ (unfavored) display a similar pattern.

\textit{Sensitivity on Fragmentation Functions---}\ignorespaces
In Fig.~\ref{fig:ff_comparison} we examine the compatibility of the ABCMO data with different modern
 FF sets.  We consider the following parametrizations: \mbox{\code{BDSSV22\_NNLO}} \cite{Borsa:2022vvp}, \code{DSS07\_NLO} \cite{deFlorian:2007aj}, \mbox{\code{MAPFF10\_NNLO}} \cite{AbdulKhalek:2022laj}, \mbox{\code{NNFF10\_NNLO}} \cite{Bertone:2017tyb}, \mbox{\code{NPC23\_NLO}} \cite{Gao:2024nkz}, and \mbox{\code{NPC23\_NNLO}} \cite{Gao:2025hlm}
and compute the hadron multiplicity distribution~\eqref{eq:RhSIDIS} at NNLO.  We verified that the multiplicity is insensitive to the choice of PDF.

Both NLO sets \cite{deFlorian:2007aj} and \cite{Gao:2024nkz} provide a good description of the data, with a tendency of overshooting the data at small $z$ for \mbox{\code{NPC23\_NLO}}.
When comparing NNLO sets instead, we observe significant differences.
Among all NNLO sets, the \mbox{\code{BDSSV22\_NNLO}} set provides the best
description of the data.
This is most likely due to the inclusion of lower-virtuality ($Q^2 < 4\, \mathrm{GeV}^2$) SIDIS data in the fit \cite{Borsa:2022vvp}.
The \mbox{\code{NPC23\_NNLO}} set provides an overall good description of the data, in particular at moderate and large $z$ for the favored channels.
The excess of \mbox{\code{NPC23\_NNLO}} in the low $z$ region for the flavor-favored processes can be attributed to the inclusion of $e^+e^-$ data from BESIII \cite{BESIII:2025mbc} into the fit.
The \mbox{\code{NNFF10\_NNLO}} set also provides an overall satisfactory description of the data, but it does not capture the shape of the distributions as accurately as the former and systematically overshoots the data by $50\,\%$ in the favored channels.
This is most likely due to the absence of SIDIS data in this fit, which help constrain the magnitude of
individual  FFs.
The \mbox{\code{MAPFF10\_NNLO}} provides a reasonable description of the data in the $\nu p \to \mu^-\pi^+ X$ channel as well as in the unfavored channels for $z\gtrsim 0.2$.  Unlike the other FF sets, the  \mbox{\code{MAPFF10\_NNLO}} FFs (central value and all replicas) fail to predict the shape and size for $\bar\nu p\to \mu^+ \pi^- X$. This is due to a significant amount of isospin breaking in this parametrization: $D^{\pi^+}_u \neq D^{\pi^-}_d$. 

We verified that the spread among the different parametrizations in Fig.~\ref{fig:ff_comparison}
 is considerably larger than the
uncertainty on each parametrization, as quantified by their associated replicas.
This spread clearly demonstrates
the stringent constraints that \mbox{(anti-)}neutrino SIDIS data provide on the flavor structure of the FFs, highlighting their
potential relevance for future global FF fits.
\begin{figure}[t]
\centering
\includegraphics[width=0.4\textwidth]{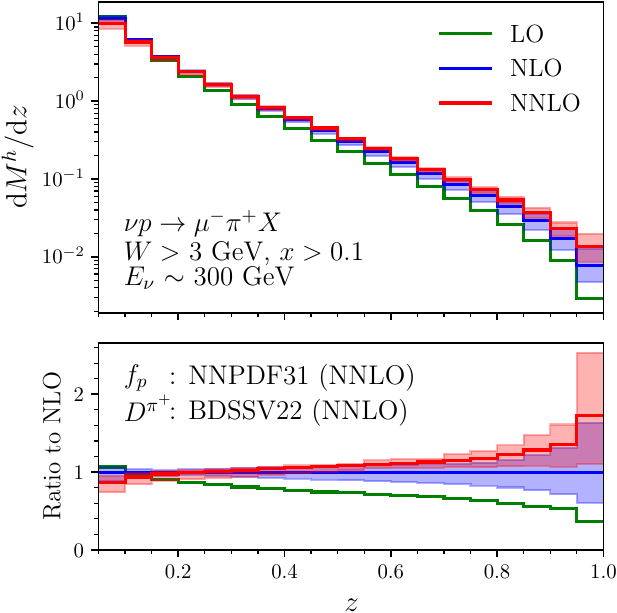}
\caption{Expected pion multiplicity distribution for FPF-type experiments at the LHC.}
\label{fig:fpf}
\end{figure}

\textit{Future Experiments---}\ignorespaces
The FASER experiment, which is situated in the very-forward region of the ATLAS interaction point at the
CERN LHC, has established the flux of forward high-energy neutrinos from proton-proton collisions~\cite{FASER:2024ref}.
Experiments in the context of the planned Forward Physics Facility (FPF,~\cite{Feng:2022inv}) at the HL-LHC could
be equipped to study SIDIS processes for various final-state hadron species.
In Fig.~\ref{fig:fpf} we investigate the behavior of the $\pi^+$ multiplicity distribution~\eqref{eq:RhSIDIS} in such a
measurement.
We consider a representative muon neutrino energy of $E_{\nu}\sim 300\, \mathrm{GeV}$ \cite{Kling:2023tgr} and employ the same kinematic cuts as before: $x>0.1$ and $W > 3\,\mathrm{GeV}$.
In the kinematic range of the FPF
 the multiplicity displays a non-negligible dependence on $E_{\nu}$
 such that more dedicated studies will require a weighted average over the neutrino beam energy.
We observe sizable QCD corrections also at higher energies with improved perturbative stability with respect to Fig.~\ref{fig:ABCMO} due to the higher average  value of $Q$:
the increase in energy results in a drastic reduction of scale uncertainties at NNLO. The corrections become
sizable only for large $z$, which can be attributed to soft gluon radiation, requiring resummation in this region.

\textit{Conclusions---}\ignorespaces
We have performed the first NNLO QCD precision study of
identified hadron production in (anti-)neutrino-induced DIS processes.
By comparing our newly derived theoretical predictions to legacy fixed-target data from
the ABCMO collaboration~\cite{Aachen-Bonn-CERN-Munich-Oxford:1982jrr},
we observe a considerably improved description of their kinematical shape upon inclusion of the NNLO
corrections. We
also demonstrate the high sensitivity of these data on the hadron FFs, thus calling for the
inclusion of formerly ignored (anti-)neutrino SIDIS data in future global determinations of FFs,
and for future measurements at the CERN FPF.

%----------------------------------------
\begin{acknowledgments}
\textit{Acknowledgments---}\ignorespaces
%----------------------------------------
We are thankful to Valerio Bertone, Ignacio Borsa, Jun Gao, Emanuele Nocera, and Rodolfo Sassot for useful discussions.
This work has received funding from the Swiss National Science Foundation (SNF)
under contract 200020-204200 and from the European Research Council (ERC) under
the European Union's Horizon 2020 research and innovation programme grant
agreement 101019620 (ERC Advanced Grant TOPUP).
\end{acknowledgments}

%----------------------------------------
\textit{Data availability---}\ignorespaces
%----------------------------------------
The data points shown in the figures are made available through~\cite{plot_data}.

%----------------------------------------
\bibliography{nuSIDIS}
%----------------------------------------

\end{document}